\documentclass[journal]{vgtc}                     

\onlineid{1971}

\vgtccategory{Research}

\title{Beyond Problem Solving: Framing and Problem–Solution Co-Evolution in Data Visualization Design}

\author{%
  \authororcid{Paul C. Parsons}{0000-0002-4179-9686},
  \authororcid{Prakash Chandra Shukla}{0009-0002-7416-1758}
}

\authorfooter{
  \item
  	Paul Parsons is with Purdue University.
  	E-mail: parsonsp@purdue.edu
  \item
  	Prakash Shukla is with Purdue University.
  	E-mail: shukla37@purdue.edu
}

\abstract{%
Visualization design is often described as a process of solving a well-defined problem by navigating a design space. While existing visualization design models have provided valuable structure and guidance, they tend to foreground technical problem-solving and underemphasize the interpretive, judgment-based aspects of design. In contrast, research in other design disciplines has emphasized the importance of framing---how designers define and redefine what the problem is---and the co-evolution of problem and solution spaces through reflective practice. These dimensions remain underexplored in visualization research, particularly from the perspective of expert practitioners. This paper investigates how visualization designers frame problems and navigate the interplay between problem understanding and solution development. We conducted a mixed-methods study with 11 expert design practitioners using design challenges, diary entries, and semi-structured interviews. Through reflexive thematic analysis, we identified key strategies that participants used to frame design problems, reframe them in response to evolving constraints or insights, and construct bridges between problem and solution spaces. These included the use of metaphors, heuristics, sketching, primary generators, and reflective evaluation of failed or incomplete ideas. Our findings contribute an empirically grounded account of visualization design as a reflective, co-evolutionary practice. We show that framing is not a preliminary step, but a continuous activity embedded in the act of designing. Participants frequently shifted their understanding of the problem based on solution attempts, feedback from tools, and ethical or narrative concerns. These insights extend current visualization design models and highlight the need for frameworks that better account for framing and interpretive judgment. We conclude with implications for visualization research, education, and practice. In particular, we discuss how design education can better support framing and co-evolutionary thinking, and how visualization research can benefit from greater attention to the cognitive strategies and reflective processes that underpin expert design.
}

\keywords{Data visualization design, data visualization practice, problem framing, problem-solution co-evolution, design cognition.}

\graphicspath{{figs/}{figures/}{pictures/}{images/}{./}} 

\usepackage{tabu}                      
\usepackage{booktabs}                  
\usepackage{lipsum}                    
\usepackage{mwe}                       

\usepackage{mathptmx}                  

\begin{document}



\maketitle

\section{Introduction}
Understanding the nature of the visualization design process has long been a central concern in visualization research. Over the past decade, several influential models and frameworks have provided valuable structure for this endeavor. These models, such as the Nested Model \cite{munzner_nested_2009}, the Design Study Methodology \cite{sedlmair_design_2012}, and the Design Activity Framework \cite{mckenna_design_2014}, offer practical guidance for navigating complex design contexts and have helped clarify how visualization designers move from understanding a problem domain to producing visual solutions. Common across many of these models is an emphasis on defining the design problem, characterizing tasks, and systematically exploring a space of potential solutions. This view often aligns with a broadly rationalist perspective in which design is seen as an informed and iterative process of optimizing solutions based on defined goals and constraints.

These frameworks have enriched our understanding of visualization design and continue to serve as important reference points for both researchers and educators. At the same time, recent work has called attention to aspects of design practice that are less emphasized in existing models---particularly the interpretive, situated, and judgment-based nature of design activity \cite{meyer_criteria_2020, parsons_understanding_2021, akbaba_entanglements_2025, dork_critical_2013, McCurdy2016, parsons_judgment_2025}. Design theorists such as Schön \cite{schon_reflective_1983} and Dorst \cite{dorst_frame_2015} have argued that design is not only about solving predefined problems, but also about constructing and redefining those problems through reflective practice. Designers often engage in framing: selecting, interpreting, and prioritizing elements of a complex situation to shape the direction of their work \cite{nelson_design_2012, schon_problems_1984}. In this view, the process of understanding what the problem is evolves in tandem with efforts to develop a solution \cite{dorst_co-evolution_2019}.

Despite its centrality in design theory, framing has received limited attention in visualization research, especially as it relates to practitioners’ own thinking and decision-making. Most existing work on framing in visualization focuses on how frames affect end-user interpretation (e.g., \cite{hullman_visualization_2011, dimara_task-based_2018, kong_frames_2018, parsons_conceptual_2018}), rather than how designers actively construct frames as part of their process. Similarly, the relationship between problem understanding and solution generation is often presented sequentially, with the problem defined up front and the solution developed thereafter. Yet studies in other design domains suggest that problem and solution spaces frequently co-evolve, with solution ideas prompting reframing of the problem, and vice versa \cite{dorst_exploring_2003, ball_advancing_2019}.

In this paper, we investigate visualization design as a reflective, co-evolutionary practice. We draw on empirical data from expert practitioners to examine how they construct and revise problem framings, and how their understanding of the problem evolves alongside emerging design ideas. Specifically, we address the following research questions:

\begin{enumerate}
    \item How do visualization design practitioners engage in problem framing, and what are its characteristics?
    \item What is the relationship between problem formulation and solution generation in data visualization practice?
\end{enumerate}

Through a mixed-methods study involving design challenges, diary entries, and interviews with expert designers, we identify a range of framing strategies, co-evolutionary patterns, and cognitive mechanisms that shape the design process. These include the use of conceptual bridges, primary generators, metaphors, constraints, and judgments to connect and reshape problem and solution spaces. Our contributions are threefold. First, we offer a detailed empirical account of how problem framing functions in expert visualization practice. Second, we provide evidence of co-evolution between problem and solution spaces, including the strategies designers use to navigate this interplay. Third, we discuss the implications of these findings for visualization theory, design education, and research methodology. In doing so, we build on and extend existing models by foregrounding the interpretive and reflective dimensions of design.

\section{Background}
Designing data visualizations often involves navigating ambiguity, complexity, and competing constraints. To support this process, researchers have developed a variety of models and frameworks to describe how designers move from an initial understanding of the problem to the development of effective visual solutions \cite{munzner_nested_2009, meyer_nested_2015, mckenna_design_2014, sedlmair_design_2012, sedig_design_2016, McCurdy2016, roberts_sketching_2016}. These models have played a foundational role in shaping both visualization theory and pedagogy. Yet as visualization challenges grow more open-ended and context-sensitive, researchers have increasingly recognized the need to better account for the interpretive, situated, and iterative nature of design. In this section, we review relevant ideas from design theory that help explain how design practitioners formulate problems and develop solutions in tandem. We focus on two core constructs: (i) the co-evolution of problem and solution spaces, and (ii) the role of framing as a central cognitive and creative activity in design.

\subsection{Design Cognition and Co-Evolution}
Traditional accounts of design cognition---the cognitive processes involved in how designers reason, make judgments, and generate ideas during design activities---often draw from the ``rational problem solving'' paradigm, which conceptualizes design as a structured search process: define the problem, identify constraints, and then systematically generate or select an optimal solution \cite{goldschmidt_linkography_2014, dorst_design_2006}. This view, strongly associated with Herbert Simon’s work \cite{simon_sciences_1969, simon_structure_1973}, describes problem-solving as transforming an initial state into a goal state through a series of logical operations. In this framework, the problem is assumed to exist independently of the designer and can be progressively structured until a solution can be found. Even when problems are seen as ``ill-structured'' at the outset, the goal is to convert them into well-structured problems that lend themselves to rational solution strategies. As Simon wrote, ``much problem-solving effort is directed at structuring problems, and only a fraction of it at solving problems once they are structured.'' \cite{simon_structure_1973} This paradigm has been deeply influential across disciplines, including HCI and visualization. However, it has also been critiqued for inadequately capturing the reality of design work, particularly when dealing with complex, ambiguous, or ``wicked'' problems. These critiques point to the limitations of assuming a linear progression from problem definition to solution generation, and they emphasize the designer’s active role in interpreting, shaping, and reframing the problem itself.

An alternative perspective is provided by Donald Schön’s theory of reflective practice \cite{schon_reflective_1983}, which describes design as a dynamic, dialogic activity. In this view, the problem is not fixed or fully knowable at the start; rather, it is constructed and reconstructed through engagement with materials, stakeholders, and emerging ideas. Schön used the concept of ``framing'' to highlight that designers often work by identifying what to pay attention to, how to understand a situation, and what direction to pursue next \cite{schon_problems_1984}. This framing process is iterative and embedded in action: designers think through making and reflect in and on their design moves \cite{schon_designing_1992}. Building on Schön’s insights, scholars such as Dorst \cite{dorst_co-evolution_2019}, Maher \cite{maher_co-evolution_2003}, and Cross \cite{dorst_creativity_2001} proposed the concept of \textit{problem–solution co-evolution} as a more accurate model of creative design activity. Rather than treating problem definition and solution development as separate stages, co-evolution describes them as mutually influential: working on a tentative solution often reveals new aspects of the problem, which in turn reshapes the direction of solution development. This interplay is seen as a hallmark of expert design behavior, particularly in domains characterized by uncertainty or ambiguity.

\subsection{Framing}
Framing in design refers to the active process by which designers shape how a situation is understood. Rather than assuming the problem is fixed, framing emphasizes that problems are interpreted through the designer’s perspective, values, and context. Drawing on Schön’s reflective practice and extended by Dorst’s model of frame creation \cite{dorst_frame_2015}, framing is not just about naming or describing the problem---it involves identifying tensions, surfacing opportunities, and making a ``creative leap'' to articulate a new conceptual lens through which the situation can be reinterpreted. This new frame acts as an organizing principle that guides the search for and evaluation of design directions. Framing is especially important when dealing with wicked problems—those that are ill-defined, value-laden, and resistant to clear solutions \cite{rittel_dilemmas_1973}. In such contexts, framing becomes a continuous and judgment-driven activity. Designers make framing judgments about what to emphasize or exclude, which stakeholders to prioritize, or which interpretive angle to pursue. These judgments shape the scope and trajectory of the project and help make sense of ambiguous goals or shifting constraints \cite{darke_primary_1979, nelson_design_2012, gray_judgment_2015}. In this work, we adopt a broad view of framing that includes both initial problem framing and the framing judgments that occur throughout the design process. We treat framing as a core interpretive activity through which designers define what matters, determine the scope of inquiry, and influence the form and function of the resulting visualization.

\subsubsection{Framing Devices: Primary Generators and Bridges}
To support our analysis of how designers move between problem framing and solution development, we draw on two complementary concepts from design theory: \textit{primary generators} and \textit{bridges}. The term \textit{primary generator} was introduced by Darke \cite{darke_primary_1979} to describe a central idea that emerges early in the design process, often before the problem is fully understood. This idea may be a metaphor, a concept, or a visual form that acts as a catalyst, helping the designer impose structure on ambiguity. It typically becomes a core organizing principle, around which both problem understanding and solution decisions are developed. Similarly, \textit{bridges} are conceptual or material tools that support movement between the problem space and the solution space throughout the process \cite{dorst_co-evolution_2019}. They may include heuristics, framing methods, metaphors, sketching techniques, or visual structures. While a primary generator often serves as the initial conceptual anchor, bridges are used more iteratively, to explore ideas, reconcile tensions, or reinterpret goals in light of emerging designs. These concepts provide a vocabulary for analyzing how designers construct meaning and make progress under uncertainty.

\section{Visualization Design Theory and Models}
Numerous models and frameworks have been proposed to characterize and guide the visualization design process. Popular examples include the Nested Model \cite{munzner_nested_2009} and its Blocks and Guidelines extension \cite{meyer_nested_2015}, the nine-stage framework in the Design Study Methodology (DSM) \cite{sedlmair_design_2012}, and the Design Activity Framework \cite{mckenna_design_2014}, among others \cite{goodwin_creative_2013, McCurdy2016, sedig_design_2016}. While these models offer valuable guidance, they also raise important questions about their underlying assumptions regarding design cognition. Many implicitly align with the rational problem-solving paradigm, in which design is viewed as a structured search through a pre-defined problem space to achieve specified goals \cite{simon_sciences_1969}. For example, DSM begins with ``learn'' and ``winnow'' stages focused on understanding the domain and identifying relevant questions before moving to solution-oriented phases like ``design'' and ``implement''. Similarly, the Nested Model structures design decisions hierarchically, starting from domain problems and task abstraction, and progressing through visual encoding choices to algorithm design. This top-down sequencing suggests a separation between problem and solution, with the former presumed to be defined in advance and used to constrain the latter.

Such approaches reflect Simon’s view that even ill-structured problems can and should be clarified early in the process to enable rational solution search. The emphasis on clearly defining user needs, data, and tasks before generating visual encodings illustrates an effort to structure wicked problems, echoing Simon’s strategy for addressing ambiguity. Likewise, many models prioritize evaluation against predefined criteria, reinforcing the assumption that problem definition precedes and guides solution generation. While some frameworks include reflection, it is typically positioned late in the process and framed as a means of contributing to academic discourse rather than as a core mechanism of design cognition \cite{parsons_design_2023}. For instance, the DSM includes ``reflection'' as its eighth stage, focused on articulating insights post hoc. Meyer and Dykes \cite{meyer_reflection_2018} have observed a ``bias towards post-study reflection'' and advocate for more intentional reflection throughout the design process.

In contrast to Simon’s model of rational problem-solving, Donald Schön’s paradigm of reflective practice emphasizes the fluid, situated nature of design work. In this view, problems and solutions co-evolve through iterative cycles of framing, action, and reflection. From this perspective, visualization design would be expected to involve ongoing judgment, imagination, and situated negotiation rather than adherence to a known path or sequence. Despite growing recognition that visualization problems are often wicked or ill-structured, most models do not fully embrace the reflective practice perspective. Simply labeling design as ``iterative'' or ``complex'' does not clarify whether a model supports reflective practice or remains rooted in a rationalist view. Furthermore, the development of many design models has not explicitly engaged with theories of design cognition, possibly due to limited awareness of this literature within the visualization community \cite{parsons_design_2023}.

Meyer and Dykes \cite{meyer_reflection_2018} argue that visualization research remains largely shaped by positivist assumptions, favoring generalizable knowledge and systematic methods. As a result, frameworks developed within this paradigm tend to prioritize clarity, abstraction, and evaluation---qualities that align with Simon’s paradigm but may marginalize the ambiguity, negotiation, and judgment emphasized in Schön’s \cite{parsons_design_2020}. This dynamic can produce a self-reinforcing cycle: researchers develop models consistent with prevailing epistemologies, which are then assessed according to the same standards that shaped their design. Finally, while these models are influential in the academic visualization community, their relationship to actual design practice remains uncertain \cite{parsons_understanding_2021}. Below is an elaboration on how positivist and rationalist views on design manifest in visualization design models and frameworks:

\textbf{Emphasis on defined inputs and outputs, aligning with the ``problem as given'': } 
Many traditional visualization design models implicitly assume that the problem to be addressed is pre-defined. Stages like ``understand the data'' or ``characterize the task'' suggest that the designer’s role is to interpret a given problem rather than question or redefine it. For example, the Nested Model begins with ``characterize the task and data,'' emphasizing translation of domain knowledge into visualization terms without addressing how the problem itself might be reframed. Similarly, the nine-stage DSM framework starts with ``learn'' and ``winnow'' to identify key domain challenges, but quickly moves toward solution-oriented stages like ``cast,'' ``design,'' and ``implement,'' with limited emphasis on revisiting the initial framing. The Design Activity Framework \cite{mckenna_design_2014} includes phases like ``understand,'' ``ideate,'' and ``make,'' which allow flexibility but still largely center on the problem as presented. While these models acknowledge iteration, they generally do not foreground the possibility of radically reframing the problem during the design process.

\textbf{Focus on the solution space:} A significant portion of visualization research and model development has historically focused on the effectiveness and efficiency of visual encodings and interaction techniques. This emphasis aligns with a problem-solving approach where the goal is to find the most effective way to represent data to solve a pre-defined analytical task. Mackinlay's early work on automated presentation tools \cite{Mackinlay1986}, which viewed design as a search for optimal graphical representations based on effectiveness and expressiveness, is a prime example of this problem-solving orientation. The nested model itself, while valuable for understanding design decisions at different levels, primarily guides the solution space once the domain tasks and data abstractions are established. It provides a framework for making and validating decisions about how to visualize, rather than explicitly addressing the initial framing of what needs to be visualized and why that problem definition is the most relevant or insightful.

\textbf{Implicit assumption of task clarity:} Some frameworks implicitly assume a level of task clarity that might not exist in real-world, ill-defined design problems. The task clarity axis proposed in the DSM \cite{sedlmair_design_2012} highlights the spectrum from fuzzy to crisp tasks. However, many earlier models and even some contemporary ones tend to focus more on how to address tasks that are at least somewhat defined, rather than providing extensive guidance on how to clarify or redefine inherently fuzzy domain tasks through problem framing.

\textbf{Limited emphasis on designer subjectivity and interpretive role:} Problem framing is a highly interpretive activity that relies on the designer's ability to bring a unique perspective to a situation. However, many visualization models, with their emphasis on systematic processes and objective evaluation criteria, may not fully capture or value this subjective aspect of design. The focus tends to be on applying established visualization principles and guidelines to a given problem, rather than on the designer's role in actively shaping the understanding of that problem.

\textbf{Post-study reflection rather than ongoing reframing:} While reflection is acknowledged as an important part of the design study methodology, it is often positioned as a stage that occurs after the design, implementation, and deployment phases. This post-study reflection is crucial for extracting lessons learned and refining guidelines, but it might not adequately capture the iterative and in-the-moment reframing that expert designers engage in throughout the design process as they gain deeper insights into the problem. Dominant perspectives have a bias towards post-study reflection \cite{meyer_criteria_2020, parsons_design_2023}, and there is a need to better integrate reframing activities within existing models.

\textbf{Influence of the rational problem-solving paradigm:} The underlying paradigm of design cognition significantly influences the assumptions embedded in visualization models. The historical dominance of the rational problem-solving view, influenced by Simon's work, has led to models that emphasize structured search, logical steps, and optimization within a defined problem space. This paradigm inherently downplays the importance of problem framing, which Schön argued is a more fundamental aspect of design, especially when dealing with ill-defined situations. Many existing visualization models and frameworks appear closer in spirit to Simon's view, supporting decision-making and rational search, rather than Schön's reflective practice view, which would highlight problem framing, judgment, and the situated nature of design \cite{parsons_design_2023}.

\section{Methods}
This study draws on the tradition of practice-focused design research, which seeks to understand design as it unfolds in real-world contexts, attending to the thoughts, actions, and judgments of practitioners in situ \cite{kuutti_turn_2014, stolterman_design_2012}. Rather than isolating specific variables under laboratory conditions, this approach prioritizes naturalistic and situated methods that can capture the interpretive, iterative, and reflective nature of design work. Our aim was to investigate how expert visualization practitioners engage in problem framing and navigate the co-evolution of problem and solution spaces---not as abstract functions, but as activities embedded in specific projects, tools, constraints, and value systems.

Practice-focused research has become increasingly important in visualization, as scholars have sought to understand the complexities of design beyond performance metrics or artifact evaluation (e.g., \cite{bako_understanding_2022, parsons_understanding_2021, bako_unveiling_2025, walny_data_2020, alspaugh_futzing_2019, stokes_its_2025, baigelenov_how_2025}). To pursue this line of inquiry, we employed a multi-method, longitudinal approach that balances generative elicitation, in-the-moment reflection, and retrospective sensemaking. Specifically, we used three complementary methods: (1) design probes to stimulate reflective activity and elicit participants’ early framing moves in a controlled but open-ended design task; (2) diary records to capture participants’ ongoing reasoning, judgments, and design decisions over the course of real projects; and (3) semi-structured interviews to retrospectively surface participants’ rationale, interpretive shifts, and insights across both probe and diary contexts.

This combination was selected to support both depth and triangulation: probes allowed us to observe early-stage framing in a shared context across participants; diaries captured longitudinal engagement with actual professional work; and interviews provided a space for clarification, elaboration, and integration of earlier reflections. Taken together, these methods enabled us to examine how framing and co-evolutionary reasoning occur across different settings, timescales, and levels of abstraction. Participants were recruited from March 2023 through February 2024, and data was collected from September 2023 through July 2024. Prior to recruitment, our study was reviewed and approved by our institution’s ethics board.

\subsection{Design Challenge}
To investigate how practitioners engage in early-stage framing and co-evolutionary design thinking, we developed a design probe centered around a shared, open-ended design challenge. The goal was to elicit reflective activity in a semi-controlled context while allowing for interpretive freedom. We selected the inaugural ``Data Is Plural'' design challenge, hosted by the Data Visualization Society (DVS) in 2022 \cite{nightingale_editors_data_2022}, as the basis for the probe. This challenge was chosen because it is publicly available, recognized as valuable within the practitioner community, does not require specialized expertise, and is designed to encourage creativity in visualization practice. Participants were given a suggested time of 90 minutes to complete the challenge independently---at a time and place of their choosing---without direct interaction with the researchers during the process. All communications about the task, including instructions and submission guidelines, occurred via email. Participants were informed via email that we were primarily interested in their thinking and process rather than the completeness or polish of the final outcome. They were encouraged to submit rough or partial artifacts, such as sketches, screenshots, annotations, or notes, as evidence of their design process.

To capture their evolving thought process, participants were asked to submit short written or recorded reflections at two key points: (i) immediately after reading the design brief and (ii) upon completing the design challenge. These checkpoints allowed us to observe how participants framed the problem at the outset, how their understanding evolved during exploration, and how they made sense of their design trajectory after completing the task. Participants were also prompted to explain their rationale for design decisions, including how they interpreted the problem and what considerations guided their solution development. Although the challenge was contrived and time-bounded, using a shared design context across all participants provided a basis for cross-case comparison while preserving the interpretive and exploratory character of professional design work. The resulting artifacts and reflections served as valuable material for analyzing early framing activity and the co-evolution of problem and solution spaces.

\subsection{Diary Entries}
To capture design activity in real-world contexts, participants were asked to maintain a reflective diary over the course of an active visualization project from their professional work. This method was intended to document how framing judgments and design decisions unfold over time, in situ, and under the constraints and dynamics of authentic practice. Diary methods are especially useful for surfacing tacit knowledge and cognitive processes that may not be easily recalled in retrospective interviews, while also preserving the situatedness of design reasoning. 

Participants received detailed instructions for the diary exercise via email following their completion of the design challenge. These instructions outlined expectations for diary entries, emphasizing reflective documentation of their ongoing professional visualization projects. Participants recorded their entries approximately once per week, with flexibility to accommodate the pace and rhythm of their particular project. Prompts encouraged them to reflect on how they were framing the problem, what kinds of judgments they were making, what design directions they were pursuing or discarding, and how external constraints (e.g., feedback, time, data limitations) shaped their process. Participants were invited to submit intermediate artifacts---such as notes, sketches, prototypes, or screenshots---accompanied by brief verbal or written annotations describing their thinking at the time of creation. The diaries provided a longitudinal view of participants’ design activity, allowing us to examine framing and co-evolution as ongoing processes rather than discrete events. By situating our analysis within the natural cadence of participants’ work, the diary entries served as a bridge between the more structured design probe and the reflective interview, offering insight into how framing behaviors manifest over the course of real projects.  

\subsection{Interviews}
Following the design probe and diary phases, each participant completed a 90-minute semi-structured interview. Interviews were conducted remotely via Zoom, led by the first author with the second author acting as a note-taker. Interviews were recorded and automatically transcribed using Dovetail \cite{noauthor_dovetail_2025}, then manually reviewed and corrected by the researchers. The interviews provided opportunities to delve deeper into the reasoning behind participants’ decisions, clarify observations from the prior phases, and reflect more broadly on their framing strategies and design process. Semi-structured interviews were selected to balance consistency across participants with the flexibility to pursue emergent insights and participant-led narratives. The interview protocol included three major parts: (i) reflections on the design probe activity, (ii) discussion of their diary project and related artifacts, and (iii) broader reflections on their approach to problem framing and design strategy across projects. Participants were asked to revisit key moments from their design probe and diary submissions—such as sketches, visual experiments, or turning points in the project—and to elaborate on the thinking that informed their choices. Where applicable, participants were prompted to describe alternative paths they considered, what they discarded and why, and how their understanding of the problem evolved over time.

We also explored participants’ general design practices, including how they define and redefine problems, manage ambiguity, make interpretive judgments, and structure their work in response to constraints. Follow-up questions encouraged participants to articulate their tacit methods, such as metaphors, heuristics, or techniques they regularly use to move between problem and solution spaces. The interview data provided a valuable complement to the design probe and diary materials, enabling triangulation and deeper interpretive insight. In particular, the interviews helped contextualize design decisions that may not have been fully explained in earlier artifacts or diary entries, and allowed participants to articulate the reflective and interpretive aspects of their work that are often difficult to capture in real-time.

\subsection{Participants}
Our study focused on expert data visualization practitioners. We excluded novice designers, as prior research suggests that experts are more likely to demonstrate the reflective, judgment-based practices that are central to our research questions \cite{christiaans_cognitive_1992, schon_reflective_1983}. Experts possess a richer repertoire of strategies for navigating ambiguity, managing competing constraints, and engaging in the co-evolution of problems and solutions, making them particularly well-suited for exploring framing and interpretive design processes. To identify expert participants, we drew on Hoffman’s Pentapod Principle \cite{hoffman_identifying_2023} and prior scholarship on design expertise \cite{atman_engineering_2007}, which emphasize expertise as multifaceted and not merely determined by years of experience. These frameworks highlight the importance of indicators such as professional accomplishments, peer recognition, and demonstrable impact in the field.

Based on these insights, we developed criteria to guide participant selection. Eligible practitioners demonstrated one or more of the following: (a) significant experience in data visualization, including leadership roles on design teams; (b) awards for visualization work (e.g., \textit{Information is Beautiful}); (c) speaking engagements at influential venues such as \textit{OpenVisConf}, \textit{Eyeo}, \textit{Outlier}, or \textit{Visualized}; (d) peer recognition through features in books, podcasts (e.g., \textit{PolicyViz}, \textit{Storytelling with Data}), or a substantial social media presence (e.g., 1,500+ followers); and (e) public exhibitions of their work. Additionally, we reviewed participants’ portfolios to assess quality, complexity, and participants' demonstrated ability to navigate ill-structured problems, work across diverse contexts, and articulate design decisions beyond surface-level execution. While no participant was expected to meet every criterion, we sought diversity in experience and perspective. Participants were recruited via multiple channels: direct outreach, the Data Visualization Society’s recruitment mechanism, and public calls on platforms like Reddit, Twitter, and LinkedIn. Each participant was compensated with \$250 USD. Table~\ref{tab:participantdemographics} provides an overview of our 11 participants, including pseudonyms, self-reported titles, years of experience, and recognized markers of expertise.
\begin{table*}
\centering
  \caption{Demographics of our study participants, including pseudonyms, professional titles, years of experience, and markers of expertise.}
  \label{tab:participantdemographics}
  \begin{tabular}{cccc}
    \toprule
    Participant & Title & Years of Experience & Expertise Indicators \\
    \midrule
    John & Data Experience Designer & 11+ & Speaker, awards, peer recognition \\
    Priya & Data Visualization Specialist & 2-5 & Awards, peer recognition \\
    Elena & Lead Visualization Designer & 6-10 & Speaker, leading datavis team  \\
    Carlos & Data Visualization Designer & 6-10 & Speaker, awards, exhibitions \\
    Omar & Senior Interactive Designer & 11+ & Speaker, peer recognition, leading datavis team \\
    Lauren & Creative Information Designer & 11+ & Speaker, awards, peer recognition, author \\
    Jade & Data Visualization Designer & 6-10 & Speaker, peer recognition\\
    Lina & Data Visualization Designer & 6-10 & Awards \\
    Sofia & Information Designer & 2-5 & Speaker, awards, exhibitions\\
    Maya & Data Visualization Specialist & 11+ & Speaker, awards, peer recognition \\
    Laila & Visual Data Journalist & 6-10 & Awards, peer recognition \\
  \bottomrule
\end{tabular}
\end{table*}

\subsection{Data Analysis}
Table~\ref{tab:studydata} summarizes the data collected from each study participant across the three components of the study: the design challenge, diary entries, and semi-structured interviews. For the design challenge, we report word counts from participants’ written or recorded responses at two key points (R1: after reading the brief; R2: after exploring the dataset). The average word count for diary entries reflects the level of engagement and depth in documenting their ongoing design work. Interview durations ranged from 72 to 101 minutes, with the average at 87 minutes, reflecting rich, in-depth conversations about participants’ reasoning, design judgments, and framing processes across both the challenge and their real-world projects.

\begin{table}[h!]
  \caption{Study participants with word counts from design challenge responses (R1, R2), average word count from diary entries, and interview durations.}
  \label{tab:studydata}
  \centering
  \begin{tabular}{cccc}
    \toprule
    Participant&Challenge&Diary Entry&Interview Duration\\
    \midrule
    John & 2706, 1245 & 231 & 1 hr 27 mins\\
    Priya & 290, 392 & 405 & 1 hr 28 mins\\
    Elena & 525, 594 & 390 & 1 hr 32 mins\\
    Carlos & 1539, 1214 & 434 & 1 hr 30 mins\\
    Omar & 2987, 2918 & 531 & 1 hr 23 mins\\
    Lauren & 453, 464 & 257 & 1 hr 19 mins\\
    Jade & 251, 459 & 366 & 1 hr 33 mins\\
    Lina & 2209, 3213 & 270 & 1 hr 35 mins\\
    Sofia & 369, 393 & 139 & 1 hr 41 mins\\
    Maya & 1164, 425 & 445 & 1 hr 22 mins\\
    Laila & 163, 447 & 238 & 1 hr 12 mins\\
  \bottomrule
\end{tabular}
\end{table}

We conducted a reflexive thematic analysis \cite{braun_using_2006} to examine participants’ reflections, artifacts, and design activities across the three stages of the study. This method was chosen for its capacity to surface patterns of meaning across qualitative data while attending to variation, complexity, and the interpretive role of the researcher. Reflexive thematic analysis recognizes that coding and theme development are not mechanical processes, but involve active, situated judgment informed by theory, context, and perspective. Our analysis followed a hybrid approach that combined both deductive and inductive strategies \cite{fereday_demonstrating_2006}. This allowed us to apply relevant concepts from design theory---such as problem framing, co-evolution, and framing judgments---while also remaining open to new and unanticipated insights in the data. We adopted the three-phase hybrid model outlined by Swain \cite{swain_hybrid_2018}, which builds on the work of Fereday and Muir-Cochrane \cite{fereday_demonstrating_2006}, and provides a structured yet flexible framework for analyzing complex design phenomena.

Both authors were involved in the analysis, with the first author coordinating the process. We began with an extended period of familiarization, reading and re-reading interview transcripts and diary entries to note preliminary observations about how participants navigated the relationship between problem understanding and design development. We then developed an initial coding scheme that included both a priori codes, drawn from existing theory, and emergent, inductively derived codes, generated through close reading of the data. These included segments where participants framed or reframed the problem, made key design decisions, responded to constraints, or transitioned between problem and solution spaces. Given the conceptual clarity of our a priori codes, application was generally straightforward, and coders reached consensus without the need for formal inter-rater reliability scores---an approach consistent with recent critiques of rigid reliability metrics in qualitative research \cite{mcdonald_reliability_2019}. The inductive coding process, though grounded in the data, was informed by our familiarity with design cognition and visualization literature. Codes and memos were organized using Dovetail, enabling collaborative refinement and traceability throughout the process.

From these codes, we iteratively developed candidate themes that represented recurring strategies, behaviors, and design judgments. Themes were reviewed and revised through team discussions, with attention to within-participant coherence and cross-participant variation. We defined each theme in relation to its explanatory value for our research questions, carefully delineating conceptual boundaries and connections. Throughout the process, we engaged in reflexive dialogue about our own assumptions and interpretive frames, recognizing that analysis is shaped by the researchers’ positionality. Our aim was not to quantify behavior, but to construct a rich, grounded account of how expert visualization designers frame problems, respond to ambiguity, and navigate the co-evolution of problem and solution spaces.

\section{Findings: Framing and Problem--Solution Co-Evolution in Visualization Design Practice}
Our analysis revealed two interrelated but distinct aspects of visualization design practice: how designers engage in problem framing, and how problem and solution spaces co-evolve through the course of a project. First, we found that problem framing was not treated as a one-time or front-loaded activity. Instead, participants actively constructed, revised, and negotiated what the problem was---often multiple times---in response to data exploration, narrative ideas, feasibility constraints, or audience considerations. Framing was situated, interpretive, and guided by judgment, rather than derived from fixed task requirements. Second, we observed that framing and solution development did not occur in a strict sequence. Instead, they frequently co-evolved: participants moved fluidly between understanding the challenge and generating design responses. Sketches, metaphors, design methods, tool feedback, and failed experiments served as bridges between these evolving spaces. The following themes highlight how participants engaged in both framing and co-evolution throughout their design processes.

\subsection{Co-evolution through Visual Experimentation}
Many participants engaged in sketching, prototyping, and visual exploration not just to refine solutions, but to better understand the problem itself. These early solution attempts acted as probes that revealed structure, ambiguity, or gaps in their understanding. As Sofia put it, \textit{``I use visualization to make mistakes''}---positioning the design process as a means of inquiry rather than execution. John echoed a similar stance: \textit{``I started making a couple of sketches \ldots not necessarily in the direction of the final visualization yet. It is rather a more visual exploration.''} Elena described how even quick visualizations helped her test viability early: \textit{``Just throwing some really high level stuff out there \ldots this was my way of seeing off the cuff if that idea is even gonna be viable.''} Lina also described sketching as a thinking tool: \textit{``Sketching that out visually is helping me answer those questions \ldots there’s an infinite number of ways I could reorganize this data.''} 

\subsection{Reframing in Response to Design Constraints}
Constraints within the solution space---such as tool limitations, time pressure, or clarity of output---often prompted participants to reframe the problem itself. Elena, for example, described a limitation in her tool: \textit{``I couldn’t figure out how to do this in Power BI \ldots so I opted for small multiples.''} Maya encountered a time constraint that shifted her approach: \textit{``I had to change gears when I realized how much time it was going to take.''} Jade similarly shifted focus after finding the actor and performance data in the design challenge overly granular, opting instead to emphasize time, theater, and performance type. Lauren described her adjustment as a shift in fidelity: \textit{``My goal was just to give you guys a couple sketches of what I might try to bring into Figma.''} 

\subsection{Conceptual Anchors and Bridges}
Participants often relied on concepts, metaphors, or framing heuristics to move fluidly between the problem and solution spaces. These anchors helped them articulate design directions and stabilize meaning. Lina grounded her framing in an architectural metaphor, drawing inspiration from the circular layout of theater spaces and audience arrangements. Lauren used a self-defined heuristic to clarify the significance of the topic and guide her framing decisions. Maya invoked the idea of marketing and social media to think about how theaters could have their own social media accounts, moving between a potential solution and reframing of the data. Omar described bridging between problem and solution as a natural mode of thinking: \textit{``Even while you are still opening, opening, opening \ldots it’s only natural that you are already thinking in those terms [of solutions].''} Maya also noted the importance of testing conceptual ideas visually: \textit{``If I was gonna make the flower work, there are a few things that it had to be able to do to function \ldots I tested it with one flower to make sure it worked.''} 

\subsection{Framing through Solution Rejection}
Some of the clearest moments of problem reframing occurred when participants abandoned a design direction that failed to convey the intended message. In these cases, rejection was not a failure, but a productive part of the process. Elena abandoned an initial area chart design after finding that the data distribution lacked the clarity and interest she had anticipated, reflecting: \textit{``I did have a specific visualization that I was curious about \ldots [after creating the visualization] it gave me the answer right away. I was like, no, the data isn't clean in that specific way \ldots [I need to] go come up with some other directions to explore.''} John began with a visualization but realized it was \textit{“too cluttered”} to reveal meaningful structure and found that \textit{“you couldn’t see anything.”} After trying \textit{“different approaches”} without success, he reframed the challenge---not as refining the existing layout, but as rethinking the structure entirely. By \textit{“rotating everything”}, he discovered a more legible pattern, which shifted his understanding of what the visualization could reveal. This reframing turned a failed design into a moment of insight: \textit{“that was actually the moment for me when I thought, OK, this is going to work}.”

\subsection{Structured Co-Evolution}
While many participants moved between problem and solution informally, others developed repeatable methods to structure this movement. These methods helped them generate framing options and evaluate alternatives. Maya described one such approach: \textit{``I do what I call the cube method---come up with six ways of doing something, then choose.''} Lauren used a cyclical method: \textit{``I repeat [my custom method] \ldots get an angle, do more exploration, then do it again.''} She added, \textit{``It was more about making sure I understood the data and a good question.''} 

\subsection{Empathy and Interpretation as Generative Forces}
Designers also drew on empathic reasoning, personal ethics, and aesthetic sensibilities to shape both framing and design decisions. Lauren grounded her framing in imagined user needs: \textit{``I'm actually trying to put myself in that place of wherever that data came from. If I were going to a show \ldots what would I care about?''} Sofia described a project involving mortality data and the importance of bringing humanity to it. She described how the initial treemaps were \textit{``all too cold and we are not putting the focus on the people that have passed away, so that's when the beeswarm plot came in.''} Laila similarly sought emotional resonance based on a source of inspiration from a different project: \textit{``And then I start thinking of a treemap \ldots what if I made it more fun like that speaker one \ldots just more artsy.''} 

\subsection{Primary Generators as Organizing Concepts}
Several participants described the emergence of an early, compelling idea or metaphor that acted as a central organizing principle for the rest of the project. These ``primary generators'' \cite{darke_primary_1979} emerged before participants had fully articulated the design problem or evaluated the complete dataset, yet they helped guide decision-making across both problem and solution spaces. Rather than being fixed solutions, these ideas shaped how the participants came to understand what their project was fundamentally about. Sofia, for instance, described her early conceptual commitment to a dandelion metaphor to visualize shared language among employees: \textit{``Each seed is a word \ldots the dandelion gets bigger depending on the number of times a specific word was used.''} She explained that the metaphor emerged not from data analysis, but from a desire to emphasize unity and conversation, offering a conceptual direction that later helped her structure the visual and narrative aspects of the design. Similarly, Priya described how the idea of gamification emerged early on as a possible way to structure the visualization around bureaucratic barriers in a cross-border data story: \textit{``From the beginning \ldots the idea of gamification was there \ldots it could make sense because it’s a journey full of obstacles.''} This framing not only influenced visual and interaction design, but also clarified narrative priorities and user perspective. Laila also described how an early sketch led her to commit to a treemap structure, which then served as a base concept to evolve visual layout and stylistic decisions: \textit{``I was looking more toward having it like a timeline \ldots then I started thinking of a treemap \ldots and then thought, what if I made it more fun like that speaker one.''} This generative idea guided both aesthetic and structural choices throughout the process. These examples illustrate how primary generators provided a nucleus around which problem framing and solution development coalesced. They were not simply outputs, but catalysts for co-evolution—helping participants impose structure on ambiguity while retaining interpretive flexibility.

\subsection{Creative Leaps in Frame Creation}
In several cases, participants made what Dorst describes as a ``creative leap''---a shift in perspective that resulted in the articulation of a new frame \cite{dorst_frame_2015}. These moments were more than iterative refinements; they involved stepping back from the initial problem formulation to reimagine the nature of the challenge. The resulting frames not only helped participants structure the design space more effectively, but also surfaced deeper questions of value, narrative, and audience.  Lauren, for example, described struggling with a tree metaphor to explain communication dynamics before arriving at a more resonant concept: the potted plant. \textit{``If you have a potted plant, then the person can take some action \ldots that made so much more sense and people don’t need to be explained to that you need to take care of a potted plant.''} This shift in metaphor reshaped how she thought about agency and responsibility, and led her to a comic-style narrative structure that better fit the diversity of user stories. 

Omar described how he reframed a technically complex project about AI agents by treating it as a children's app even though they would not be the only users: \textit{``my strategy for the agents project has been framing it as a kids app, even if 80\% of the project is really building a framework for agents that operate with data \ldots the strategy pays off in different ways, especially keeping the design clean and super simple, avoiding technicalities, assuming the user is intelligent and creative but doesn't have a background in data science or coding \ldots at the end of the day, all these benefit any kind of user.''} This strategic reframing simplified the design language and helped him align more closely with core user needs. 


\subsection{Vignettes of Framing and Co-Evolution in Practice}
To further illustrate how problem framing and solution development co-evolve, we present three vignettes offering detailed views of individual participants' design processes. They were selected for richness and clarity in showing how participants moved between interpreting the problem and exploring solutions. Each vignette highlights different framing mechanisms---including metaphor, values, structural methods, and reframing in response to ambiguity---while demonstrating the reflective and situated nature of visualization practice.

\subsubsection{Lauren: Framing Through Metaphor and Structure}
Lauren’s design process offers a vivid example of how framing and solution development co-evolve through cycles of reflection, metaphor, and structured experimentation. Early in her project, she grappled with how to convey interpersonal dynamics within a communication dataset. Her initial approach used a tree metaphor to map branching roles, but it felt overly hierarchical and failed to capture the shared, relational qualities she wanted to express. In response, Lauren made a conceptual shift, reframing the problem around actionability and care. This led to a new metaphor: the potted plant. This frame—a living, low-maintenance object that requires attention to thrive—offered a more resonant way to visualize interpersonal responsibility. It grounded her work in familiar, emotionally meaningful imagery and helped her clarify the type of message she wanted to convey. The metaphor wasn’t just visual; it reoriented her understanding of the problem itself. Instead of simply describing relationships, the visualization would prompt the viewer to consider what it means to nurture them.

Lauren’s process also included a structured framing heuristic she developed herself, focused on identifying context, surfacing tensions, and developing resolutions. She used this heuristic iteratively to explore different angles, assess whether her visual choices clarified central conflicts, and refine her narrative direction. Through these cycles, her framing became progressively more grounded in ethical and empathic design priorities. Eventually, Lauren adopted a comic-strip layout, pairing the plant metaphor with narrative panels. This format allowed her to show not only the state of the data but also the actions people might take in response—further reinforcing her goal of design for engagement, not just representation. This vignette illustrates how conceptual framing, metaphor, and heuristics can structure a co-evolutionary design process. Lauren didn’t merely iterate on visuals—she used them as tools to think through what the problem actually was. Her case exemplifies the interpretive flexibility and framing judgment that our findings highlight as core to expert visualization practice.

\subsubsection{Omar: A Primary Generator for Simplicity}
Omar’s project began with a goal of setting a strategy for education on data literacy and AI. This defined a problem space and gave the project an initial direction. Omar adopted a “kids as personas” approach, which functioned as a primary generator, translating the broad problem into a concrete, initial design approach. The strategy immediately provided a clear heuristic for generating potential solutions by prompting Omar to repeatedly ask whether \textit{``a kid cares about this or that?''} This framing led directly to potential characteristics of the solution: prioritizing simplicity, avoiding technicalities, and clear UI elements. The solution strategy became a lens for interpreting what the product should do and how it should feel. 

Yet this solution also generated a new problem, as Omar described wondering: \textit{``maybe I’m going too far with the ‘kids strategy’ \ldots will these adult users be affected, maybe underwhelmed, by the ‘childish’ UI?”} The very success of the approach in clarifying the design raised new concerns about how the interface might be perceived by adult users. Omar now faced a new framing challenge---how to retain the simplicity and clarity afforded by the child-centered design without alienating broader audiences. Rather than following a linear trajectory from problem to solution, Omar’s process exemplifies co-evolution. The kids-focused solution did not merely respond to a fixed problem; it continually redefined what the problem was, what it could become, and what kinds of solutions were appropriate. This dynamic interplay between framing, design action, and emerging understanding was central to Omar’s process.

\subsubsection{Lina: Reframing toward Advocacy-Centered Design}
Lina’s vignette offers a clear example of value-driven framing in the face of ambiguity. Lina was hired to contribute to a civic technology project with the goal of building a dashboard that \textit{``shows polling locations and equity.''} However, the concept of equity was, in her words, “extremely vague” and “fairly ill-defined” by the client. Although the client had some ideas, such as highlighting racial disparities or time to poll, the product’s specific goals remained unclear. Rather than proceeding with the vague brief, Lina reframed the problem: instead of creating a generic data display, she envisioned a tool that would directly support the advocacy work of legal organizations and community groups. Her starting point was to ask, \textit{“Who are our partners? What are their goals? What are their pain points?”} This shift redefined the fundamental purpose of the application---from showing data to assisting users in real-world workflows, such as identifying where to open new polling locations.

Lina’s initial user interviews and card-sorting exercises, even when conducted with an organization that was “not” truly user-focused, were attempts to co-evolve the problem and the solution. By asking users to rank potential features, she sought to uncover what the product “might be useful” for—refining the problem definition through explorations of possible solutions. She also questioned core assumptions about the solution form. In particular, she challenged the client’s default reliance on a map interface. Through sketching alternative designs, she asked, \textit{“Is a map the correct thing?”} and \textit{“Is it really answering the questions that they want answered?”} This illustrates a continuous re-evaluation of solution forms against the evolving understanding of the problem—even though external pressures ultimately led to adopting a map-based approach.

In summary, Lina consistently demonstrates the ability to reframe ill-defined problem briefs, uncover underlying user needs, and interrogate dominant solution assumptions. Her framing work actively shapes the trajectory of design, enabling the problem and solution to co-evolve through iterative inquiry and reflective critique. By actively shaping what the problem was about—and for whom—she redefined the scope of the project and foregrounded dimensions of meaning that might otherwise have remained invisible.

\subsubsection{Commentary: Design Narratives, Theoretical Threads}
While these vignettes offer illustrative stories, they also concretize key aspects of our theoretical framing. Lauren’s narrative highlights the role of framing judgments and conceptual metaphors in structuring a design trajectory, showing how solution ideas can actively reshape the designer’s understanding of the problem. Omar’s account provides a compelling example of a creative reframing that redefined not just visual aesthetics, but the underlying goals of the project exemplifying a ``creative leap.'' Lina’s vignette foregrounds the ethical and values-based dimensions of framing, reminding us that problem definition in visualization is never neutral. Her refusal of superficial metrics and use of spatial metaphor exemplify the designer’s interpretive agency.

Together, these vignettes demonstrate how visualization designers do not merely navigate a predefined space; rather, they construct and revise the space itself. They move fluidly between exploration and articulation, between sensing ambiguity and generating form. These cases reaffirm our argument that problem framing is not a preliminary phase, but a central and ongoing activity within visualization design. They also underscore the value of attending to designers’ own conceptual tools—such as metaphors, heuristics, and narrative devices—as key mechanisms for bridging problem and solution spaces.

\section{Discussion}
Our findings reveal that problem--solution co-evolution is not an incidental phenomenon, but a fundamental feature of how visualization designers work, particularly in open-ended or exploratory contexts. This challenges traditional assumptions that design begins with a clearly defined problem and proceeds toward a solution. Instead, we observed designers navigating an evolving interplay between understanding what the problem is and shaping what the solution could be. This interplay was often triggered by friction: unclear data structure, unsatisfying early visualizations, infeasible ideas, or interpretive ambiguity. These frictions did not stall progress; rather, they provoked reframing. What began as a question about performance counts became a story about theater infrastructure (Elena); what began as a heat map became a treemap narrative of theatrical scale (Laila). These shifts show that solution attempts are not endpoints, but probes---tools for inquiry that reshape the designer’s understanding of what matters.

Participants also employed a range of framing tools that functioned as bridges across problem and solution spaces. Some were visual (sketches, Tableau tabs, layout experiments), some conceptual (metaphors, heuristics), and others procedural (tool constraints, scoping boundaries, iterative ideation). These bridges enabled designers to stabilize the relationship between evolving understanding and emerging form. Rather than selecting from a pre-defined design space, participants created the terms of that space through exploration. Several participants---including Lauren, Maya, and Omar---demonstrated a high degree of self-awareness about their methods, often reflecting on their reasoning mid-process. This suggests that co-evolutionary design is not only common but can be articulated and potentially taught. For example, Maya’s cube method or Lauren’s framing heuristic could be introduced as pedagogical scaffolds for helping students productively alternate between exploring problem meaning and testing design ideas.

The designers in our study did not simply `solve' visualization problems---they actively constructed them. Their design work was inseparable from their interpretive work. This supports a view of visualization as not only technical, but also conceptual and editorial. Design tools like Tableau, Illustrator, and sketchpads became part of a broader epistemic process: a way of knowing what the problem is, not just a means to present what has already been defined. Taken together, these findings deepen our understanding of visualization design practice by offering a more granular account of how designers navigate ambiguity and progress through their work. Moreover, they highlight the importance of fostering environments---both educational and professional---that allow for ambiguity, encourage exploration, and support iterative framing as a core component of visualization thinking.

\subsection{Expanding Visualization Design Theory}
Our findings contribute to ongoing conversations in the visualization literature about how theory and models can more fully reflect the realities of design practice. In this section, we draw connections between our empirical results and existing theoretical frameworks, highlighting where our study complements, extends, or complicates current understandings. We focus in particular on how framing and problem–solution co-evolution can deepen our grasp of iteration, illuminate the interpretive role of designers, and suggest opportunities for advancing future models of visualization design.

\subsubsection{Complementing Existing Models of Visualization Design}
Our findings deepen our understanding of visualization design practice by offering a more granular account of how designers navigate ambiguity and progress through their work. Models such as the Nested Model \cite{munzner_nested_2009}, the Design Study Methodology \cite{sedlmair_design_2012}, and the Design Activity Framework \cite{mckenna_design_2014} have made important contributions to understanding visualization design as a flexible, non-linear, and iterative process. Our study illuminates the interpretive and co-evolutionary mechanisms that animate this process in practice, highlighting framing, reframing, and conceptual bridging as central drivers of progress and meaning-making.

\subsubsection{Extending the Account of Framing in Visualization Design}
While many existing frameworks implicitly include or allow for framing in early stages (e.g., task characterization, domain scoping), our findings show that framing is not confined to the beginning of the process. Instead, it often continues throughout design, with participants frequently revisiting or reframing their understanding of the problem in response to what they learn from attempted solutions. For example, solution attempts that failed to clarify patterns (as in Elena’s case) or visual experiments that revealed unexpected structures (as with Omar or Priya) led participants to reshape their problem definitions. This supports and extends recent calls in the field \cite{meyer_reflection_2018, parsons_design_2023} for a richer, more reflective approach to framing and iteration.

\subsubsection{Supporting Work That Recenters the Designer’s Role}
Our findings also align with a growing body of visualization research that emphasizes the designer’s role as an interpretive agent \cite{dork_critical_2013, offenhuber_autographic_2023, parsons_understanding_2021, akbaba_entanglements_2025, dignazio_data_2023}. These perspectives highlight how visualization is shaped by situated judgment, political context, and epistemic framing. Our study complements these perspectives by showing how such judgment manifests moment-to-moment during the design process, and how design tools and visual experiments become sites of sensemaking, not just production. By capturing how participants navigated uncertainty, reframed problems, and used design moves to surface meaning, this work helps bring design cognition into clearer focus within the visualization literature.

\subsubsection{Informing Future Theory and Models}
In light of these findings, we believe that existing theory and models might benefit from an expanded vocabulary for describing the co-evolutionary aspects of design. Rather than representing framing as a discrete early phase, future models could incorporate it as an ongoing, revisable activity. More importantly, framing should be acknowledged not merely as a variation of ``understanding'' or ``discovering,'' but as an \textit{active, interpretive stance} that designers take throughout the design process. This interpretive stance influences how designers define what matters, prioritize competing concerns, and negotiate meaning through engagement with data, stakeholders, and tools.

\subsection{Implications for Research}
This work contributes to the growing body of scholarship that views data visualization as a reflective design practice. By highlighting the interpretive, framing-based, and co-evolutionary nature of expert practice, our findings offer several implications for researchers studying visualization design. First, this work encourages researchers to move beyond models that treat visualization as primarily a technical or procedural activity. Instead of focusing solely on decisions made within a predefined design space, scholars should attend to how that space is constructed in the first place—how practitioners define what the problem is, what matters, and what should be emphasized or excluded. This perspective aligns with and extends recent calls to account for designer agency, judgment, and situated reasoning in visualization research \cite{parsons_understanding_2021, dork_critical_2013, offenhuber_autographic_2023}.

Second, this study provides a vocabulary for describing specific cognitive strategies that designers use to navigate complexity. Concepts such as framing judgments, bridges, and primary generators offer analytic tools that researchers can use to characterize the trajectory of a project, or to interpret turning points within a design process. These constructs may help researchers analyze practice in richer ways, moving beyond checklists of activities or retrospective evaluations of final artifacts. Future work might explore how often these mechanisms occur in various design settings, how they relate to project constraints, or how they influence downstream outcomes. Third, this work has methodological implications. Researchers who aim to study design practice should consider collecting data that captures moments of interpretive shift and reasoning in action—such as sketches, discarded concepts, reflections during prototyping, and design rationales. Our use of design probes, diary methods, and interviews illustrates one possible approach. As prior work has shown, studying practice on its own terms often requires open-ended, longitudinal, and participant-driven methods \cite{kuutti_turn_2014, walny_data_2020, bako_understanding_2022}.

Finally, this study supports and extends practice-focused research agendas in visualization. By foregrounding framing and co-evolution, it challenges prevailing assumptions about linearity, objectivity, and task clarity in design. It also opens new opportunities for comparative studies of how framing behaviors differ across domains, expertise levels, or team settings. Researchers interested in shaping future design methods, tools, or educational approaches may find it valuable to foreground the kinds of framing and judgment-based activities that our participants demonstrated throughout their processes. This study also contributes to expanding the conceptual language available to design researchers. Building on prior work in visualization design cognition \cite{parsons_design_2023}, we identify specific mechanisms—such as primary generators and bridges—that help explain how design trajectories evolve in response to emerging understanding. These concepts may support finer-grained analysis of design behavior and offer shared vocabulary for studying and teaching reflective practice in visualization.

\subsection{Implications for Education}

Our findings have implications for how students are prepared to navigate real-world, ill-structured design challenges. Current educational approaches often emphasize technical fluency and adherence to established design models. Our study suggests that visualization practice involves substantial interpretive work that is not well captured by existing models. In particular, problem framing and the co-evolution of problem and solution spaces are not merely advanced skills---they are foundational to how expert designers operate. This calls for a rethinking of educational priorities and instructional approaches.

Work in various fields has emphasized the need to cultivate design judgment as a central learning outcome in design education \cite{parsons_developing_2023, korkmaz_development_2014, demiral-uzan_instructional_2024}. Our findings extend this by highlighting specific forms of judgment—especially framing judgments—that are central to expert practice. As Nelson and Stolterman argue, framing is the ``passkey to the overall formation of the design palette'' \cite{nelson_design_2012}. Educators should therefore consider how curricula can help students develop the ability to define problems, question assumptions, and articulate their framing rationale throughout the design process. Rather than treating problem framing as a front-end exercise (e.g., during task analysis), instruction should support students in revisiting and refining their frames as their design understanding evolves.

To that end, visualization pedagogy may benefit from scaffolding co-evolutionary thinking. As observed in this study, many expert practitioners used heuristics or structured methods (e.g., Maya’s cube method, Lauren’s framing heuristic) to move between framing and designing. These methods could be integrated into studio courses or project-based assignments to help students experiment with framing as an iterative, generative activity. Recent work in the context of UX Design education \cite{parsons_developing_2023} shows that students perceive framing judgment ability as central to their growth and identity as designers, and that exposure to reflective, exploratory, and dialogic teaching methods plays a key role in developing this capacity. Educators can draw on these insights to create classroom environments that encourage ambiguity, inquiry, and conceptual risk-taking.

Several researchers have noted the challenges of teaching visualization in ways that balance structure with open-endedness \cite{bach_challenges_2023, adar_roboviz_2023}. Our findings underscore that such ambiguity is not only pedagogically valuable, but necessary if students are to engage in authentic framing and problem-setting. This suggests a shift from strictly performance-based evaluations of design outcomes to assessments that consider how students frame problems, justify decisions, and demonstrate reflective awareness throughout the process. Drawing inspiration from research on judgment development in HCI, instructional design, and engineering \cite{parsons_developing_2023, korkmaz_development_2014, gray_revealing_2019, beckman_teaching_2012}, instructors might provide formative feedback on framing decisions, incorporate design critiques focused on problem construction, or embed reflection prompts that encourage students to articulate how their understanding of the problem has evolved.

In sum, visualization education should move beyond emphasizing solutions and design artifacts to also foreground the interpretive and judgment-based processes that precede and guide those artifacts. Doing so may better prepare students for the complexities and uncertainties of visualization practice, and help them cultivate the framing and co-evolutionary capabilities that are essential to expert design work.

\subsection{Limitations}
While this study offers new insights into visualization design practice, several limitations should be acknowledged. First, the participant sample, while diverse in background and experience, was limited to 11 individuals working within a specific design challenge context. This context was intentionally exploratory and open-ended, which may have encouraged framing and co-evolution behaviors more than task-driven or production-oriented environments typically do. However, as discussed in the section above, participants also described similar framing activities in their professional practice, providing reassurance that these behaviors are not artifacts of the study setup. Second, the data collection relied on a combination of self-reported diary entries, design artifacts, and interviews. While this multi-modal approach helped triangulate findings, it also depended on participants' ability and willingness to reflect on and articulate their thought processes. Some framing judgments or shifts in understanding may have occurred tacitly and gone unreported.

Third, the study emphasizes depth over breadth. Our goal was not to quantify how often specific behaviors occur, but to qualitatively explore the nature of framing and co-evolution in detail. As a result, the findings are not meant to be statistically generalizable but instead serve to surface patterns, vocabulary, and mechanisms that can inform future studies. Finally, while we identified a range of framing behaviors and co-evolutionary patterns, we did not systematically evaluate their outcomes in terms of visualization quality, user impact, or effectiveness. Future work could build on this foundation by linking framing practices to downstream design outcomes and reception. Despite these limitations, the study provides a grounded and detailed account of how framing and co-evolution unfold in visualization design, offering both conceptual and methodological contributions for researchers and educators in the field.

\section{Conclusion}
This study examined how expert visualization designers engage in problem framing and the co-evolution of problem and solution spaces. Through design probes, diary records, and interviews with practitioners, we identified a range of interpretive and iterative strategies that designers use to make sense of complex, ambiguous design contexts. Rather than treating problem formulation as a preliminary or fixed step, participants continuously shaped and reshaped their understanding of the problem throughout the design process. Their framing activities were influenced not only by technical constraints or user needs, but also by personal values, ethical considerations, and aesthetic judgments. Our findings challenge traditional representations of visualization design as a rational, stage-based process and call attention to the reflective, situated nature of design cognition. Designers in our study did not simply apply expertise within a predefined design space; they actively constructed and revised the terms of that space, using sketches, metaphors, heuristics, and tool feedback to explore and articulate the problem in tandem with emerging design ideas. This supports a view of visualization design as a form of reflective practice, in which framing is not a one-time event but a continuous, generative activity.

By surfacing the mechanisms through which designers navigate ambiguity—such as conceptual bridges, primary generators, and iterative reframing—we offer both theoretical and practical contributions. For researchers, this work extends existing models by highlighting the need to study design not only as a matter of solution effectiveness, but also as a process of meaning-making and judgment. For educators, our findings suggest the value of teaching framing and co-evolution as foundational capabilities, not merely advanced skills. Scaffolding these practices can better prepare students to engage with the open-ended, ill-defined challenges that characterize much of real-world visualization work. Ultimately, this study contributes to an evolving discourse that views visualization not only as a technical craft, but as a complex design practice grounded in interpretation, inquiry, and reflection. By foregrounding the interpretive and judgment-based aspects of visualization work, we hope to support a richer understanding of what it means to design visualizations—and to inform future research, pedagogy, and practice in the field.

\acknowledgments{
We wish to thank the study participants for their valuable contributions. This work was supported by NSF grant \#2146228.}

\bibliographystyle{abbrv-doi-hyperref}

\bibliography{references}

\begin{thebibliography}{10}

\bibitem{noauthor_dovetail_2025}
Dovetail, 2025.
\newblock https://dovetail.com/.

\bibitem{adar_roboviz_2023}
E.~Adar and E.~Lee-Robbins.
\newblock Roboviz: {A} {Game}-{Centered} {Project} for {Information} {Visualization} {Education}.
\newblock {\em IEEE Transactions on Visualization and Computer Graphics}, 29(1):268--277, Jan. 2023. \href{https://doi.org/10.1109/TVCG.2022.3209402}
{doi: {{%
10\hspace{.1pt}\discretionary{.}{%
}{.}\hspace{.4pt}1109\discretionary{/}{%
}{/}TVCG\hspace{.1pt}\discretionary{.}{%
}{.}\hspace{.4pt}2022\hspace{.1pt}\discretionary{.}{%
}{.}\hspace{.4pt}3209402}}}


\bibitem{akbaba_entanglements_2025}
D.~Akbaba, L.~Klein, and M.~Meyer.
\newblock Entanglements for {Visualization}: {Changing} {Research} {Outcomes} through {Feminist} {Theory}.
\newblock {\em IEEE Transactions on Visualization and Computer Graphics}, 31(1):1279--1289, Jan. 2025. \href{https://doi.org/10.1109/TVCG.2024.3456171}
{doi: {{%
10\hspace{.1pt}\discretionary{.}{%
}{.}\hspace{.4pt}1109\discretionary{/}{%
}{/}TVCG\hspace{.1pt}\discretionary{.}{%
}{.}\hspace{.4pt}2024\hspace{.1pt}\discretionary{.}{%
}{.}\hspace{.4pt}3456171}}}


\bibitem{alspaugh_futzing_2019}
S.~Alspaugh, N.~Zokaei, A.~Liu, C.~Jin, and M.~A. Hearst.
\newblock Futzing and {Moseying}: {Interviews} with {Professional} {Data} {Analysts} on {Exploration} {Practices}.
\newblock {\em IEEE Transactions on Visualization and Computer Graphics}, 25(1):22--31, Jan. 2019. \href{https://doi.org/10.1109/TVCG.2018.2865040}
{doi: {{%
10\hspace{.1pt}\discretionary{.}{%
}{.}\hspace{.4pt}1109\discretionary{/}{%
}{/}TVCG\hspace{.1pt}\discretionary{.}{%
}{.}\hspace{.4pt}2018\hspace{.1pt}\discretionary{.}{%
}{.}\hspace{.4pt}2865040}}}


\bibitem{atman_engineering_2007}
C.~J. Atman, R.~S. Adams, M.~E. Cardella, J.~Turns, S.~Mosborg, and J.~Saleem.
\newblock Engineering design processes: {A} comparison of students and expert practitioners.
\newblock {\em Journal of Engineering Education}, 96(4):359--379, 2007. \href{https://doi.org/10.1002/j.2168-9830.2007.tb00945.x}
{doi: {{%
10\hspace{.1pt}\discretionary{.}{%
}{.}\hspace{.4pt}1002\discretionary{/}{%
}{/}j\hspace{.1pt}\discretionary{.}{%
}{.}\hspace{.4pt}2168\discretionary{%
}{-}{-}9830\hspace{.1pt}\discretionary{.}{%
}{.}\hspace{.4pt}2007\hspace{.1pt}\discretionary{.}{%
}{.}\hspace{.4pt}tb00945\hspace{.1pt}\discretionary{.}{%
}{.}\hspace{.4pt}x}}}


\bibitem{bach_challenges_2023}
B.~Bach, M.~Keck, F.~Rajabiyazdi, T.~Losev, I.~Meirelles, J.~Dykes, R.~S. Laramee, M.~AlKadi, C.~Stoiber, S.~Huron, C.~Perin, L.~Morais, W.~Aigner, D.~Kosminsky, M.~Boucher, S.~Knudsen, A.~Manataki, J.~Aerts, U.~Hinrichs, J.~C. Roberts, and S.~Carpendale.
\newblock Challenges and {Opportunities} in {Data} {Visualization} {Education}: {A} {Call} to {Action}.
\newblock {\em IEEE Transactions on Visualization and Computer Graphics}, pp. 1--12, 2023. \href{https://doi.org/10.1109/TVCG.2023.3327378}
{doi: {{%
10\hspace{.1pt}\discretionary{.}{%
}{.}\hspace{.4pt}1109\discretionary{/}{%
}{/}TVCG\hspace{.1pt}\discretionary{.}{%
}{.}\hspace{.4pt}2023\hspace{.1pt}\discretionary{.}{%
}{.}\hspace{.4pt}3327378}}}


\bibitem{baigelenov_how_2025}
A.~Baigelenov, P.~Shukla, and P.~Parsons.
\newblock How {Visualization} {Designers} {Perceive} and {Use} {Inspiration}.
\newblock In {\em Proceedings of the 2025 {CHI} {Conference} on {Human} {Factors} in {Computing} {Systems}}, {CHI} '25, pp. 1--13. Association for Computing Machinery, New York, NY, USA, Apr. 2025. \href{https://doi.org/10.1145/3706598.3714191}
{doi: {{%
10\hspace{.1pt}\discretionary{.}{%
}{.}\hspace{.4pt}1145\discretionary{/}{%
}{/}3706598\hspace{.1pt}\discretionary{.}{%
}{.}\hspace{.4pt}3714191}}}


\bibitem{bako_understanding_2022}
H.~K. Bako, X.~Liu, L.~Battle, and Z.~Liu.
\newblock Understanding {How} {Designers} {Find} and {Use} {Data} {Visualization} {Examples}.
\newblock {\em IEEE Transactions on Visualization and Computer Graphics}, pp. 1--11, 2022. \href{https://doi.org/10.1109/TVCG.2022.3209490}
{doi: {{%
10\hspace{.1pt}\discretionary{.}{%
}{.}\hspace{.4pt}1109\discretionary{/}{%
}{/}TVCG\hspace{.1pt}\discretionary{.}{%
}{.}\hspace{.4pt}2022\hspace{.1pt}\discretionary{.}{%
}{.}\hspace{.4pt}3209490}}}


\bibitem{bako_unveiling_2025}
H.~K. Bako, X.~Liu, G.~Ko, H.~Song, L.~Battle, and Z.~Liu.
\newblock Unveiling {How} {Examples} {Shape} {Visualization} {Design} {Outcomes}.
\newblock {\em IEEE Transactions on Visualization and Computer Graphics}, 31(1):1137--1147, 2025. \href{https://doi.org/10.1109/TVCG.2024.3456407}
{doi: {{%
10\hspace{.1pt}\discretionary{.}{%
}{.}\hspace{.4pt}1109\discretionary{/}{%
}{/}TVCG\hspace{.1pt}\discretionary{.}{%
}{.}\hspace{.4pt}2024\hspace{.1pt}\discretionary{.}{%
}{.}\hspace{.4pt}3456407}}}


\bibitem{ball_advancing_2019}
L.~J. Ball and B.~T. Christensen.
\newblock Advancing an understanding of design cognition and design metacognition: {Progress} and prospects.
\newblock {\em Design Studies}, 65:35--59, Nov. 2019. \href{https://doi.org/10.1016/j.destud.2019.10.003}
{doi: {{%
10\hspace{.1pt}\discretionary{.}{%
}{.}\hspace{.4pt}1016\discretionary{/}{%
}{/}j\hspace{.1pt}\discretionary{.}{%
}{.}\hspace{.4pt}destud\hspace{.1pt}\discretionary{.}{%
}{.}\hspace{.4pt}2019\hspace{.1pt}\discretionary{.}{%
}{.}\hspace{.4pt}10\hspace{.1pt}\discretionary{.}{%
}{.}\hspace{.4pt}003}}}


\bibitem{beckman_teaching_2012}
S.~L. Beckman, M.~Barry, and H.~Rittel.
\newblock Teaching {Students} {Problem} {Framing} {Skills} with a {Storytelling} {Metaphor}.
\newblock {\em International Journal of Engineering Education}, 28(2):364--373, 2012.

\bibitem{braun_using_2006}
V.~Braun and V.~Clarke.
\newblock Using thematic analysis in psychology.
\newblock {\em Qualitative Research in Psychology}, 3(2):77--101, Jan. 2006. \href{https://doi.org/10.1191/1478088706qp063oa}
{doi: {{%
10\hspace{.1pt}\discretionary{.}{%
}{.}\hspace{.4pt}1191\discretionary{/}{%
}{/}1478088706qp063oa}}}


\bibitem{christiaans_cognitive_1992}
H.~Christiaans and K.~Dorst.
\newblock Cognitive {Models} in {Industrial} {Design} {Engineering}: a protocol study.
\newblock In {\em Design {Theory} and {Methodology} - {DTM92}}, pp. 131--140, 1992.

\bibitem{darke_primary_1979}
J.~Darke.
\newblock The primary generator and the design process.
\newblock {\em Design Studies}, 1(1):36--44, July 1979. \href{https://doi.org/10.1016/0142-694X(79)90027-9}
{doi: {{%
10\hspace{.1pt}\discretionary{.}{%
}{.}\hspace{.4pt}1016\discretionary{/}{%
}{/}0142\discretionary{%
}{-}{-}694X\discretionary{%
}{(}{(}79\discretionary{)}{%
}{)}90027\discretionary{%
}{-}{-}9}}}


\bibitem{demiral-uzan_instructional_2024}
M.~Demiral-Uzan and E.~Boling.
\newblock Instructional design students’ design judgment development.
\newblock {\em Educational Technology Research and Development}, 72(3):1813--1849, June 2024. \href{https://doi.org/10.1007/s11423-024-10361-1}
{doi: {{%
10\hspace{.1pt}\discretionary{.}{%
}{.}\hspace{.4pt}1007\discretionary{/}{%
}{/}s11423\discretionary{%
}{-}{-}024\discretionary{%
}{-}{-}10361\discretionary{%
}{-}{-}1}}}


\bibitem{dignazio_data_2023}
C.~D'ignazio and L.~F. Klein.
\newblock {\em Data {Feminism}}.
\newblock MIT press, 2023.

\bibitem{dimara_task-based_2018}
E.~Dimara, S.~Franconeri, C.~Plaisant, A.~Bezerianos, and P.~Dragicevic.
\newblock A task-based taxonomy of cognitive biases for information visualization.
\newblock {\em IEEE Rransactions on Visualization and Computer Graphics}, 26(2):1413--1432, 2018. \href{https://doi.org/10.1109/TVCG.2018.2872577}
{doi: {{%
10\hspace{.1pt}\discretionary{.}{%
}{.}\hspace{.4pt}1109\discretionary{/}{%
}{/}TVCG\hspace{.1pt}\discretionary{.}{%
}{.}\hspace{.4pt}2018\hspace{.1pt}\discretionary{.}{%
}{.}\hspace{.4pt}2872577}}}


\bibitem{dorst_exploring_2003}
K.~Dorst.
\newblock Exploring the {Structure} of {Design} {Problems}.
\newblock In {\em International {Conference} on {Engineering} {Design} ({ICED})}, pp. 585--588. Stockholm, 2003.

\bibitem{dorst_design_2006}
K.~Dorst.
\newblock Design {Problems} and {Design} {Paradoxes}.
\newblock {\em Design Issues}, 22(3):4--17, July 2006. \href{https://doi.org/10.1162/desi.2006.22.3.4}
{doi: {{%
10\hspace{.1pt}\discretionary{.}{%
}{.}\hspace{.4pt}1162\discretionary{/}{%
}{/}desi\hspace{.1pt}\discretionary{.}{%
}{.}\hspace{.4pt}2006\hspace{.1pt}\discretionary{.}{%
}{.}\hspace{.4pt}22\hspace{.1pt}\discretionary{.}{%
}{.}\hspace{.4pt}3\hspace{.1pt}\discretionary{.}{%
}{.}\hspace{.4pt}4}}}


\bibitem{dorst_frame_2015}
K.~Dorst.
\newblock {\em Frame {Innovation}: {Create} {New} {Thinking} by {Design}}.
\newblock Design {Thinking}. MIT Press, Apr. 2015.

\bibitem{dorst_co-evolution_2019}
K.~Dorst.
\newblock Co-evolution and emergence in design.
\newblock {\em Design Studies}, 65:60--77, Nov. 2019. \href{https://doi.org/10.1016/j.destud.2019.10.005}
{doi: {{%
10\hspace{.1pt}\discretionary{.}{%
}{.}\hspace{.4pt}1016\discretionary{/}{%
}{/}j\hspace{.1pt}\discretionary{.}{%
}{.}\hspace{.4pt}destud\hspace{.1pt}\discretionary{.}{%
}{.}\hspace{.4pt}2019\hspace{.1pt}\discretionary{.}{%
}{.}\hspace{.4pt}10\hspace{.1pt}\discretionary{.}{%
}{.}\hspace{.4pt}005}}}


\bibitem{dorst_creativity_2001}
K.~Dorst and N.~Cross.
\newblock Creativity in the design process: co-evolution of problem–solution.
\newblock {\em Design Studies}, 22(5):425--437, Sept. 2001. \href{https://doi.org/10.1016/S0142-694X(01)00009-6}
{doi: {{%
10\hspace{.1pt}\discretionary{.}{%
}{.}\hspace{.4pt}1016\discretionary{/}{%
}{/}S0142\discretionary{%
}{-}{-}694X\discretionary{%
}{(}{(}01\discretionary{)}{%
}{)}00009\discretionary{%
}{-}{-}6}}}


\bibitem{dork_critical_2013}
M.~Dörk, P.~Feng, C.~Collins, and S.~Carpendale.
\newblock Critical {InfoVis}: exploring the politics of visualization.
\newblock In {\em {CHI} '13 {Extended} {Abstracts} on {Human} {Factors} in {Computing} {Systems}}, {CHI} {EA} '13, pp. 2189--2198. Association for Computing Machinery, New York, NY, USA, Apr. 2013. \href{https://doi.org/10.1145/2468356.2468739}
{doi: {{%
10\hspace{.1pt}\discretionary{.}{%
}{.}\hspace{.4pt}1145\discretionary{/}{%
}{/}2468356\hspace{.1pt}\discretionary{.}{%
}{.}\hspace{.4pt}2468739}}}


\bibitem{nightingale_editors_data_2022}
N.~Editors.
\newblock Data {Is} {Plural} {Submissions}: {London} {Stage} {Database}, Dec. 2022.

\bibitem{fereday_demonstrating_2006}
J.~Fereday and E.~Muir-Cochrane.
\newblock Demonstrating rigor using thematic analysis: {A} hybrid approach of inductive and deductive coding and theme development.
\newblock {\em International Journal of Qualitative Methods}, 5(1):80--92, 2006.
\newblock Publisher: SAGE Publications Sage CA: Los Angeles, CA. \href{https://doi.org/10.1177/160940690600500107}
{doi: {{%
10\hspace{.1pt}\discretionary{.}{%
}{.}\hspace{.4pt}1177\discretionary{/}{%
}{/}160940690600500107}}}


\bibitem{goldschmidt_linkography_2014}
G.~Goldschmidt.
\newblock {\em Linkography: {Unfolding} the {Design} {Process}}.
\newblock MIT Press, 2014.
\newblock doi: 10.7551/mitpress/9455.001.0001.

\bibitem{goodwin_creative_2013}
S.~Goodwin, J.~Dykes, S.~Jones, I.~Dillingham, G.~Dove, A.~Duffy, A.~Kachkaev, A.~Slingsby, and J.~Wood.
\newblock Creative {User}-{Centered} {Visualization} {Design} for {Energy} {Analysts} and {Modelers}.
\newblock {\em IEEE Transactions on Visualization and Computer Graphics}, 19(12):2516--2525, 2013. \href{https://doi.org/10.1109/TVCG.2013.145}
{doi: {{%
10\hspace{.1pt}\discretionary{.}{%
}{.}\hspace{.4pt}1109\discretionary{/}{%
}{/}TVCG\hspace{.1pt}\discretionary{.}{%
}{.}\hspace{.4pt}2013\hspace{.1pt}\discretionary{.}{%
}{.}\hspace{.4pt}145}}}


\bibitem{gray_revealing_2019}
C.~M. Gray.
\newblock Revealing {Students}’ {Ethical} {Awareness} during {Problem} {Framing}.
\newblock {\em International Journal of Art \& Design Education}, 38(2):299--313, May 2019. \href{https://doi.org/10.1111/jade.12190}
{doi: {{%
10\hspace{.1pt}\discretionary{.}{%
}{.}\hspace{.4pt}1111\discretionary{/}{%
}{/}jade\hspace{.1pt}\discretionary{.}{%
}{.}\hspace{.4pt}12190}}}


\bibitem{gray_judgment_2015}
C.~M. Gray, C.~Dagli, M.~Demiral-Uzan, F.~Ergulec, V.~Tan, A.~A. Altuwaijri, K.~Gyabak, M.~Hilligoss, R.~Kizilboga, K.~Tomita, and E.~Boling.
\newblock Judgment and {Instructional} {Design}: {How} {ID} {Practitioners} {Work} {In} {Practice}.
\newblock {\em Performance Improvement Quarterly}, 28(3):25--49, 2015. \href{https://doi.org/10.1002/piq.21198}
{doi: {{%
10\hspace{.1pt}\discretionary{.}{%
}{.}\hspace{.4pt}1002\discretionary{/}{%
}{/}piq\hspace{.1pt}\discretionary{.}{%
}{.}\hspace{.4pt}21198}}}


\bibitem{hoffman_identifying_2023}
R.~R. Hoffman.
\newblock Identifying {Experts} for the {Design} of {Human}-{Centered} {Systems}: {The} {Pentapod} {Principle}.
\newblock {\em Journal of Expertise}, 6(3):259--266, 2023.

\bibitem{hullman_visualization_2011}
J.~Hullman and N.~Diakopoulos.
\newblock Visualization {Rhetoric}: {Framing} {Effects} in {Narrative} {Visualization}.
\newblock {\em IEEE Transactions on Visualization and Computer Graphics}, 17(12):2231--2240, Dec. 2011. \href{https://doi.org/10.1109/TVCG.2011.255}
{doi: {{%
10\hspace{.1pt}\discretionary{.}{%
}{.}\hspace{.4pt}1109\discretionary{/}{%
}{/}TVCG\hspace{.1pt}\discretionary{.}{%
}{.}\hspace{.4pt}2011\hspace{.1pt}\discretionary{.}{%
}{.}\hspace{.4pt}255}}}


\bibitem{kong_frames_2018}
H.-K. Kong, Z.~Liu, and K.~Karahalios.
\newblock Frames and slants in titles of visualizations on controversial topics.
\newblock In {\em Proceedings of the 2018 {CHI} {Conference} on {Human} {Factors} in {Computing} {Systems}}, pp. 1--12, 2018. \href{https://doi.org/10.1145/3173574.317401}
{doi: {{%
10\hspace{.1pt}\discretionary{.}{%
}{.}\hspace{.4pt}1145\discretionary{/}{%
}{/}3173574\hspace{.1pt}\discretionary{.}{%
}{.}\hspace{.4pt}317401}}}


\bibitem{korkmaz_development_2014}
N.~Korkmaz and E.~Boling.
\newblock Development of {Design} {Judgment} in {Instructional} {Design}: {Perspectives} from {Instructors}, {Students}, and {Instructional} {Designers}.
\newblock In B.~Hokanson and A.~Gibbons, eds., {\em Design in {Educational} {Technology}: {Design} {Thinking}, {Design} {Process}, and the {Design} {Studio}}, pp. 161--184. Springer International Publishing, Cham, 2014. \href{https://doi.org/10.1007/978-3-319-00927-8_10}
{doi: {{%
10\hspace{.1pt}\discretionary{.}{%
}{.}\hspace{.4pt}1007\discretionary{/}{%
}{/}978\discretionary{%
}{-}{-}3\discretionary{%
}{-}{-}319\discretionary{%
}{-}{-}00927\discretionary{%
}{-}{-}8\_10}}}


\bibitem{kuutti_turn_2014}
K.~Kuutti and L.~J. Bannon.
\newblock The turn to practice in {HCI}: towards a research agenda.
\newblock In {\em Proceedings of the 32nd {ACM} {Conference} on {Human} {Factors} in {Computing} {Systems}}, pp. 3543--3552. ACM Press, Toronto, Ontario, Canada, 2014. \href{https://doi.org/10.1145/2556288.2557111}
{doi: {{%
10\hspace{.1pt}\discretionary{.}{%
}{.}\hspace{.4pt}1145\discretionary{/}{%
}{/}2556288\hspace{.1pt}\discretionary{.}{%
}{.}\hspace{.4pt}2557111}}}


\bibitem{Mackinlay1986}
J.~D. Mackinlay.
\newblock Automating the design of graphical presentations of relational information.
\newblock {\em ACM Transactions on Graphics}, 5(2):110--141, 1986. \href{https://doi.org/10.1145/22949.22950}
{doi: {{%
10\hspace{.1pt}\discretionary{.}{%
}{.}\hspace{.4pt}1145\discretionary{/}{%
}{/}22949\hspace{.1pt}\discretionary{.}{%
}{.}\hspace{.4pt}22950}}}


\bibitem{maher_co-evolution_2003}
M.~Maher and H.-H. Tang.
\newblock Co-evolution as a computational and cognitive model of design.
\newblock {\em Research in Engineering Design}, 14(1):47--64, Feb. 2003. \href{https://doi.org/10.1007/s00163-002-0016-y}
{doi: {{%
10\hspace{.1pt}\discretionary{.}{%
}{.}\hspace{.4pt}1007\discretionary{/}{%
}{/}s00163\discretionary{%
}{-}{-}002\discretionary{%
}{-}{-}0016\discretionary{%
}{-}{-}y}}}


\bibitem{McCurdy2016}
N.~McCurdy, J.~Dykes, and M.~Meyer.
\newblock Action {Design} {Research} and {Visualization} {Design}.
\newblock In {\em Proceedings of the 6th {Biannual} {Workshop} on evaluation and {BEyond} - {methodoLogIcal} approaches for {Visualization} ({BELIV})}, pp. 10--18. ACM Press, New York, New York, USA, 2016. \href{https://doi.org/10.1145/2993901.2993916}
{doi: {{%
10\hspace{.1pt}\discretionary{.}{%
}{.}\hspace{.4pt}1145\discretionary{/}{%
}{/}2993901\hspace{.1pt}\discretionary{.}{%
}{.}\hspace{.4pt}2993916}}}


\bibitem{mcdonald_reliability_2019}
N.~Mcdonald, S.~Schoenebeck, and A.~Forte.
\newblock Reliability and {Inter}-rater {Reliability} in {Qualitative} {Research}: {Norms} and {Guidelines} for {CSCW} and {HCI} {Practice}.
\newblock {\em Proceedings of the ACM on Human-Computer Interaction}, 3:1--23, 2019. \href{https://doi.org/10.1145/3359174}
{doi: {{%
10\hspace{.1pt}\discretionary{.}{%
}{.}\hspace{.4pt}1145\discretionary{/}{%
}{/}3359174}}}


\bibitem{mckenna_design_2014}
S.~McKenna, D.~Mazur, J.~Agutter, and M.~Meyer.
\newblock Design {Activity} {Framework} for {Visualization} {Design}.
\newblock {\em IEEE Transactions on Visualization and Computer Graphics}, 20(12):2191--2200, 2014. \href{https://doi.org/10.1109/TVCG.2014.2346331}
{doi: {{%
10\hspace{.1pt}\discretionary{.}{%
}{.}\hspace{.4pt}1109\discretionary{/}{%
}{/}TVCG\hspace{.1pt}\discretionary{.}{%
}{.}\hspace{.4pt}2014\hspace{.1pt}\discretionary{.}{%
}{.}\hspace{.4pt}2346331}}}


\bibitem{meyer_criteria_2020}
M.~Meyer and J.~Dykes.
\newblock Criteria for {Rigor} in {Visualization} {Design} {Study}.
\newblock {\em IEEE Transactions on Visualization and Computer Graphics}, 26(1):87--97, 2020. \href{https://doi.org/10.1109/TVCG.2019.2934539}
{doi: {{%
10\hspace{.1pt}\discretionary{.}{%
}{.}\hspace{.4pt}1109\discretionary{/}{%
}{/}TVCG\hspace{.1pt}\discretionary{.}{%
}{.}\hspace{.4pt}2019\hspace{.1pt}\discretionary{.}{%
}{.}\hspace{.4pt}2934539}}}


\bibitem{meyer_reflection_2018}
M.~Meyer, J.~Dykes, and M.~Tory.
\newblock Reflection on {Reflection} in {Applied} {Visualization} {Research}.
\newblock {\em IEEE Computer Graphics and Applications}, 38(6):9--16, 2018. \href{https://doi.org/10.1109/MCG.2018.2874523}
{doi: {{%
10\hspace{.1pt}\discretionary{.}{%
}{.}\hspace{.4pt}1109\discretionary{/}{%
}{/}MCG\hspace{.1pt}\discretionary{.}{%
}{.}\hspace{.4pt}2018\hspace{.1pt}\discretionary{.}{%
}{.}\hspace{.4pt}2874523}}}


\bibitem{meyer_nested_2015}
M.~Meyer, M.~Sedlmair, P.~S. Quinan, and T.~Munzner.
\newblock The nested blocks and guidelines model.
\newblock {\em Information Visualization}, 14(3):234--249, 2015. \href{https://doi.org/10.1177/1473871613510429}
{doi: {{%
10\hspace{.1pt}\discretionary{.}{%
}{.}\hspace{.4pt}1177\discretionary{/}{%
}{/}1473871613510429}}}


\bibitem{munzner_nested_2009}
T.~Munzner.
\newblock A nested model for visualization design and validation.
\newblock In {\em {IEEE} {Transactions} on {Visualization} and {Computer} {Graphics}}, vol.~15, pp. 921--928, 2009. \href{https://doi.org/10.1109/TVCG.2009.111}
{doi: {{%
10\hspace{.1pt}\discretionary{.}{%
}{.}\hspace{.4pt}1109\discretionary{/}{%
}{/}TVCG\hspace{.1pt}\discretionary{.}{%
}{.}\hspace{.4pt}2009\hspace{.1pt}\discretionary{.}{%
}{.}\hspace{.4pt}111}}}


\bibitem{nelson_design_2012}
H.~G. Nelson and E.~Stolterman.
\newblock {\em The {Design} {Way}: {Intentional} {Change} in an {Unpredictable} {World}}.
\newblock The MIT Press, Cambridge, Massachusestts; London, England, second ed., 2012.

\bibitem{offenhuber_autographic_2023}
D.~Offenhuber.
\newblock {\em Autographic {Design}: {The} {Matter} of {Data} in a {Self}-{Inscribing} {World}}.
\newblock MIT Press, Dec. 2023.

\bibitem{parsons_conceptual_2018}
P.~Parsons.
\newblock Conceptual {Metaphor} {Theory} as a {Foundation} for {Communicative} {Visualization} {Design}.
\newblock In {\em {IEEE} {VIS} {Workshop} on {Visualization} for {Communication} ({VisComm})}, p.~6. Berlin, Germany, 2018.

\bibitem{parsons_understanding_2021}
P.~Parsons.
\newblock Understanding {Data} {Visualization} {Design} {Practice}.
\newblock {\em IEEE Transactions on Visualization and Computer Graphics}, 28(1):665--675, 2021. \href{https://doi.org/10.1109/TVCG.2021.3114959}
{doi: {{%
10\hspace{.1pt}\discretionary{.}{%
}{.}\hspace{.4pt}1109\discretionary{/}{%
}{/}TVCG\hspace{.1pt}\discretionary{.}{%
}{.}\hspace{.4pt}2021\hspace{.1pt}\discretionary{.}{%
}{.}\hspace{.4pt}3114959}}}


\bibitem{parsons_design_2023}
P.~Parsons.
\newblock Design {Cognition} in {Data} {Visualization}.
\newblock In D.~Albers~Szafir, R.~Borgo, M.~Chen, D.~J. Edwards, B.~Fisher, and L.~M.~K. Padilla, eds., {\em Visualization {Psychology}}. Springer, 2023.

\bibitem{parsons_design_2020}
P.~Parsons, C.~M. Gray, A.~Baigelenov, and I.~Carr.
\newblock Design {Judgment} in {Data} {Visualization} {Practice}.
\newblock In {\em {IEEE} {Visualization} {Conference} ({VIS})}, pp. 176--180. Salt Lake City, UT, Sept. 2020. \href{https://doi.org/10.1109/VIS47514.2020.00042}
{doi: {{%
10\hspace{.1pt}\discretionary{.}{%
}{.}\hspace{.4pt}1109\discretionary{/}{%
}{/}VIS47514\hspace{.1pt}\discretionary{.}{%
}{.}\hspace{.4pt}2020\hspace{.1pt}\discretionary{.}{%
}{.}\hspace{.4pt}00042}}}


\bibitem{parsons_judgment_2025}
P.~C. Parsons and A.~Ridley.
\newblock Judgment as {Coordination}: {A} {Joint} {Systems} {View} of {Visualization} {Design} {Practice}.
\newblock In {\em {IEEE} {Conference} on {Visualization} ({VIS})}, 2025.
\newblock arXiv:2507.01209 [cs]. \href{https://doi.org/10.48550/arXiv.2507.01209}
{doi: {{%
10\hspace{.1pt}\discretionary{.}{%
}{.}\hspace{.4pt}48550\discretionary{/}{%
}{/}arXiv\hspace{.1pt}\discretionary{.}{%
}{.}\hspace{.4pt}2507\hspace{.1pt}\discretionary{.}{%
}{.}\hspace{.4pt}01209}}}


\bibitem{parsons_developing_2023}
P.~C. Parsons, P.~C. Shukla, A.~Baigelenov, and C.~M. Gray.
\newblock Developing {Framing} {Judgment} {Ability}: {Student} {Perceptions} from a {Graduate} {UX} {Design} {Program}.
\newblock In {\em Proceedings of the 5th {Annual} {Symposium} on {HCI} {Education}}, pp. 23--32. ACM, Hamburg Germany, Apr. 2023. \href{https://doi.org/10.1145/3587399.3587401}
{doi: {{%
10\hspace{.1pt}\discretionary{.}{%
}{.}\hspace{.4pt}1145\discretionary{/}{%
}{/}3587399\hspace{.1pt}\discretionary{.}{%
}{.}\hspace{.4pt}3587401}}}


\bibitem{rittel_dilemmas_1973}
H.~W.~J. Rittel and M.~M. Webber.
\newblock Dilemmas in a general theory of planning.
\newblock {\em Policy Sciences}, 4:155--169, 1973. \href{https://doi.org/10.1007/BF01405730}
{doi: {{%
10\hspace{.1pt}\discretionary{.}{%
}{.}\hspace{.4pt}1007\discretionary{/}{%
}{/}BF01405730}}}


\bibitem{roberts_sketching_2016}
J.~C. Roberts, C.~Headleand, and P.~D. Ritsos.
\newblock Sketching {Designs} {Using} the {Five} {Design}-{Sheet} {Methodology}.
\newblock {\em IEEE Transactions on Visualization and Computer Graphics}, 22(1):419--428, 2016. \href{https://doi.org/10.1109/TVCG.2015.2467271}
{doi: {{%
10\hspace{.1pt}\discretionary{.}{%
}{.}\hspace{.4pt}1109\discretionary{/}{%
}{/}TVCG\hspace{.1pt}\discretionary{.}{%
}{.}\hspace{.4pt}2015\hspace{.1pt}\discretionary{.}{%
}{.}\hspace{.4pt}2467271}}}


\bibitem{schon_reflective_1983}
D.~A. Schön.
\newblock {\em The {Reflective} {Practitioner}: {How} {Professionals} {Think} {In} {Action}}.
\newblock Basic Books, 1983.

\bibitem{schon_problems_1984}
D.~A. Schön.
\newblock Problems, frames and perspectives on designing.
\newblock {\em Design Studies}, 5(3):132--136, 1984. \href{https://doi.org/10.1016/0142-694X(84)90002-4}
{doi: {{%
10\hspace{.1pt}\discretionary{.}{%
}{.}\hspace{.4pt}1016\discretionary{/}{%
}{/}0142\discretionary{%
}{-}{-}694X\discretionary{%
}{(}{(}84\discretionary{)}{%
}{)}90002\discretionary{%
}{-}{-}4}}}


\bibitem{schon_designing_1992}
D.~A. Schön.
\newblock Designing as a {Reflective} {Conversation} with the {Materials} of a {Design} {Situation}.
\newblock {\em Research and Engineering Design}, 3:131--147, 1992. \href{https://doi.org/10.1007/BF01580516}
{doi: {{%
10\hspace{.1pt}\discretionary{.}{%
}{.}\hspace{.4pt}1007\discretionary{/}{%
}{/}BF01580516}}}


\bibitem{sedig_design_2016}
K.~Sedig and P.~Parsons.
\newblock {\em Design of {Visualizations} for {Human}-{Information} {Interaction}: {A} {Pattern}-{Based} {Framework}}, vol.~4 of {\em Synthesis {Lectures} on {Visualization}}.
\newblock Morgan \& Claypool Publishers, Apr. 2016.
\newblock ISSN: 2159-516X. \href{https://doi.org/10.2200/S00685ED1V01Y201512VIS005}
{doi: {{%
10\hspace{.1pt}\discretionary{.}{%
}{.}\hspace{.4pt}2200\discretionary{/}{%
}{/}S00685ED1V01Y201512VIS005}}}


\bibitem{sedlmair_design_2012}
M.~Sedlmair, M.~Meyer, and T.~Munzner.
\newblock Design {Study} {Methodology}: {Reflections} from the {Trenches} and the {Stacks}.
\newblock {\em IEEE Transactions on Visualization and Computer Graphics}, 18(12):2431--2440, 2012. \href{https://doi.org/10.1109/TVCG.2012.213}
{doi: {{%
10\hspace{.1pt}\discretionary{.}{%
}{.}\hspace{.4pt}1109\discretionary{/}{%
}{/}TVCG\hspace{.1pt}\discretionary{.}{%
}{.}\hspace{.4pt}2012\hspace{.1pt}\discretionary{.}{%
}{.}\hspace{.4pt}213}}}


\bibitem{simon_sciences_1969}
H.~A. Simon.
\newblock {\em The {Sciences} of the {Artificial}}.
\newblock MIT Press, 1969.

\bibitem{simon_structure_1973}
H.~A. Simon.
\newblock The structure of ill structured problems.
\newblock {\em Artificial Intelligence}, 4(3):181--201, Dec. 1973. \href{https://doi.org/10.1016/0004-3702(73)90011-8}
{doi: {{%
10\hspace{.1pt}\discretionary{.}{%
}{.}\hspace{.4pt}1016\discretionary{/}{%
}{/}0004\discretionary{%
}{-}{-}3702\discretionary{%
}{(}{(}73\discretionary{)}{%
}{)}90011\discretionary{%
}{-}{-}8}}}


\bibitem{stokes_its_2025}
C.~Stokes, C.~Hu, and M.~A. Hearst.
\newblock “{It}'s a {Good} {Idea} to {Put} {It} {Into} {Words}”: {Writing} ‘{Rudders}’ in the {Initial} {Stages} of {Visualization} {Design}.
\newblock {\em IEEE Transactions on Visualization and Computer Graphics}, 31(1):1126--1136, Jan. 2025. \href{https://doi.org/10.1109/TVCG.2024.3456324}
{doi: {{%
10\hspace{.1pt}\discretionary{.}{%
}{.}\hspace{.4pt}1109\discretionary{/}{%
}{/}TVCG\hspace{.1pt}\discretionary{.}{%
}{.}\hspace{.4pt}2024\hspace{.1pt}\discretionary{.}{%
}{.}\hspace{.4pt}3456324}}}


\bibitem{stolterman_design_2012}
E.~Stolterman and J.~Pierce.
\newblock Design tools in practice: studying the designer-tool relationship in interaction design.
\newblock In {\em Proceedings of the {Designing} {Interactive} {Systems} {Conference} on - {DIS} '12}, p.~25. ACM Press, Newcastle Upon Tyne, United Kingdom, 2012. \href{https://doi.org/10.1145/2317956.2317961}
{doi: {{%
10\hspace{.1pt}\discretionary{.}{%
}{.}\hspace{.4pt}1145\discretionary{/}{%
}{/}2317956\hspace{.1pt}\discretionary{.}{%
}{.}\hspace{.4pt}2317961}}}


\bibitem{swain_hybrid_2018}
J.~Swain.
\newblock A hybrid approach to thematic analysis in qualitative research: {Using} a practical example.
\newblock In {\em Sage {Research} {Methods}}, vol. Part 2 of {\em Cases}. SAGE Publications, Ltd., 2018.

\bibitem{walny_data_2020}
J.~Walny, C.~Frisson, M.~West, D.~Kosminsky, S.~Knudsen, S.~Carpendale, and W.~Willett.
\newblock Data {Changes} {Everything}: {Challenges} and {Opportunities} in {Data} {Visualization} {Design} {Handoff}.
\newblock {\em IEEE Transactions on Visualization and Computer Graphics}, 26(1):12--22, Jan. 2020. \href{https://doi.org/10.1109/TVCG.2019.2934538}
{doi: {{%
10\hspace{.1pt}\discretionary{.}{%
}{.}\hspace{.4pt}1109\discretionary{/}{%
}{/}TVCG\hspace{.1pt}\discretionary{.}{%
}{.}\hspace{.4pt}2019\hspace{.1pt}\discretionary{.}{%
}{.}\hspace{.4pt}2934538}}}


\end{thebibliography}

\appendix 

\end{document}